\documentclass[journal]{IEEEtran}

\ifCLASSINFOpdf
  \usepackage[pdftex]{graphicx}
\else
  \usepackage[dvips]{graphicx}
\fi
\usepackage[cmex10]{amsmath}
\usepackage{algorithmic}

\usepackage{array}

\ifCLASSOPTIONcompsoc
\usepackage[caption=false,font=normalsize,labelfont=sf,textfont=sf]{subfig}
\else
\usepackage[caption=false,font=footnotesize]{subfig}
\fi
\usepackage{fixltx2e}
\usepackage{amsfonts}
\usepackage{amssymb}
\usepackage{amsthm}
\usepackage{bm}
\usepackage[ruled]{algorithm2e}
\usepackage[utf8]{inputenc}
\usepackage{booktabs}
\usepackage{breqn}
\usepackage{url}
\usepackage{cite}

\newcounter{tempEquationCounter} 
\newcounter{thisEquationNumber}

\renewenvironment{IEEEbiography}[1]
  {\IEEEbiographynophoto{#1}}
  {\endIEEEbiographynophoto}

\begin{document}
%
\title{A Green Perspective on Wi-Fi Offloading}

%
%
%

\author{
        Adnan~Aijaz,~\IEEEmembership{Member,~IEEE}
        and~A.~Hamid~Aghvami,~\IEEEmembership{Fellow,~IEEE}
\thanks{The authors are with the Centre for Telecommunications Research, King's College London, London, WC2R 2LS, UK. Contact email: adnan.aijaz@kcl.ac.uk}}

%
%

\markboth{IEEE Wireless Communications - Accepted for Publication}%
{Shell \MakeLowercase{\textit{et al.}}: Bare Demo of IEEEtran.cls for Journals}
%



\maketitle

\begin{abstract}
Due to increased energy consumption and carbon emissions of the ICT industry, operators worldwide are focusing on reducing energy consumption of their networks from financial as well as corporate responsibility perspectives. The subject of green or energy efficient operation of cellular access network has attracted a lot of attention in the research community recently. In this regard, dynamically powering down the radio network equipment has emerged as a promising solution. In literature, research around this type of techniques has mainly focused on quantifying the energy saving potential. However, little efforts have been made towards practical realization of these energy saving concepts. On the other hand, Wi-Fi networks are undergoing a paradigm shift towards ubiquity.  The main objective of this paper is to provide novel mechanisms for practically realizing the concept of improving power efficiency in the cellular access network through opportunistically offloading users to Wi-Fi networks. These mechanisms are based on the principles of cognitive network management. Performance evaluation shows the potential of proposed mechanisms as a viable solution for achieving energy efficiency in the cellular access network. 
\end{abstract}


%
\IEEEpeerreviewmaketitle

\section{Introduction}
%
%
%
%
\IEEEPARstart{T}{he} cellular industry has witnessed a tremendous growth over the last decade in terms of the number of subscribers and the volume of carried traffic. In order to meet the ever growing capacity requirements, operators worldwide are increasing the base station deployment density. According to one of the surveys, the number of base stations worldwide doubled between 2007 -- 2012, reaching beyond 4 million today \cite{ee_bs_sleep}.  Studies show that the Information and Communication Technology (ICT) industry, of which the cellular networks constitute a significant component, is responsible for approximately 10\% of global energy consumption, as of 2013 \cite{EC_energy}. For a cellular operator, nearly 60\% of energy is used in the access network as the base station comes out to be the most energy hungry component of the mobile network \cite{gr_commag}. Moreover, the carbon footprint of the ICT industry is expected to double to 4\% by 2020 \cite{carbon_ICT,ee_bs_sleep}.


With the increase in fuel prices worldwide, the awareness of harmful effects of $\text{CO}_2$ emissions on environment, and the depletion of non-renewable energy sources, there is a growing trend towards energy efficient or \emph{green} communications \cite{GC_CRC}. The basic objective of green communications, in a cellular sense, is to mitigate the inefficiencies in cellular network operation particularly in the access network. For cellular operators, reducing energy consumption is not just a matter of corporate responsibility, but also very much an economically important issue. 


In literature, the concept of dynamically powering down the radio network equipment has emerged as a promising solution for achieving energy efficiency in the cellular access network. Reducing the operational power consumption of base stations through sleep modes or more generally reducing the transmission power can have a significant impact on the overall power consumption of the operator in running its network. Recently, a number of studies have investigated energy efficient cellular network operation through various base station switch-off techniques.

On the other hand, the next generation of wireless communication networks is focusing on integration of wireless local area networks (e.g., Wi-Fi networks) with cellular networks (e.g., UMTS, HSPA, LTE, etc.) to utilize the combined benefits of both
technologies. Wi-Fi networks provide high data rates with limited coverage and mobility, whereas cellular networks offer comparatively low data rates but with high coverage and mobility. Much work has been done in the area of interworking between Wi-Fi and cellular networks.  There are two architectures for coupling Wi-Fi and cellular networks: \emph{loose} coupling and \emph{tight} coupling. In loose coupling architectures, the networks are independent, requiring no major co-operation between them. Service continuity is provided by roaming between the two networks. On the other hand, in a tightly coupled system the networks share a common core, and the majority of network functions such as vertical handover, resource management, billing and security are controlled and managed centrally. Nowadays, Wi-Fi is undergoing a paradigm shift towards ubiquity and outdoor/city-wide Wi-Fi networks are increasingly gaining popularity \cite{aijaz_wcm}.

In the light of above observations, we recently investigated the concept of reducing energy consumption in the cellular access network through opportunistic reallocation of users or traffic loads to Wi-Fi networks \cite{aijaz_icc_12}. The energy savings have been achieved by dynamically powering down the radio network equipment by either switching off the base station completely or removing its sectorization. Powering down of radio network equipment is extremely promising as it implies guaranteed `from-the-socket' savings. These concepts have also been investigated  in cellular to cellular offloading scenarios \cite{oliver_commag, pimrc_13, aijaz_crc}. However, like other studies, our investigation in \cite{aijaz_icc_12} mainly focussed on quantifying the energy saving potential. In literature, little efforts have been made towards practical realization of such energy saving concepts. 

Against this background, the main objective of this paper is to develop novel mechanisms for energy savings in cellular access network  through opportunistic reallocation of users to Wi-Fi networks. Realization of such energy saving mechanisms requires the management of spatio temporal network dynamics. An efficient way to handle such dynamics is to embed cognition in the network management \cite{cognition}. To this end, the key contributions of this paper are summarized as follows.

\begin{itemize}
\item We develop novel mechanisms for energy savings in the cellular access network which are based on the principles of cognitive network management. The proposed mechanisms provide two independent capabilities: a) capability of dynamically powering down the radio network equipment,  and b) capability of energy aware network selection. 

\item We introduce two Wi-Fi offloading techniques based on IP Flow Mobility \cite{IFOM}. These techniques play an integral part in the proposed energy saving mechanisms through user-centric and network-centric Wi-Fi offloading. 

\item We conduct a comprehensive performance evaluation of the proposed mechanisms in realistic multi-cellular scenarios. 
\end{itemize}


We begin our discussion by covering the preliminaries on opportunistic reallocation of users  to Wi-Fi networks and the principles of cognitive network management. Then, we introduce the proposed mechanisms which are presented in the form of two independent capabilities. After that we discuss two different Wi-Fi offloading solutions based on IP Flow Mobility. This is followed by performance evaluation of the proposed mechanisms. Finally, we conclude the article.

\section{Opportunistic Load Management for Energy Efficiency}
The concept of opportunistic reallocation of users or traffic loads from cellular to Wi-Fi networks in order to achieve energy efficiency in the cellular access network has been investigated in  \cite{aijaz_icc_12}. The energy savings are achieved by dynamically powering down the radio network equipment especially at times of low load. The reduction in traffic load in some parts of a cellular network at some times occurs due to a number of effects such as the typical day-night behavior of users, daily swarming of users from residential to corporate areas and back, and the movement of users to/from some areas at weekend and vacation times, etc.

There are two possibilities for dynamically powering down the radio network equipment: (i) turning off the base stations entirely in the cellular network at a specific time/location through traffic being sufficiently carried by another network/frequency and (ii) removing sectorization for the base stations, e.g., using spare capacity of another network/frequency to cover the required drop in load of the cellular network in order to
enable the latter to operate in omni-directional mode instead of tri-sectorized mode. Henceforth, these two techniques are referred to as the \emph{powering down} and the \emph{sectorization switching} solutions, and can be employed together in sectorized networks. Note that the radio coverage of a base station is the same in both omni-directional and tri-sectorized modes.

\begin{figure*}
\centering
\includegraphics [scale=0.57] {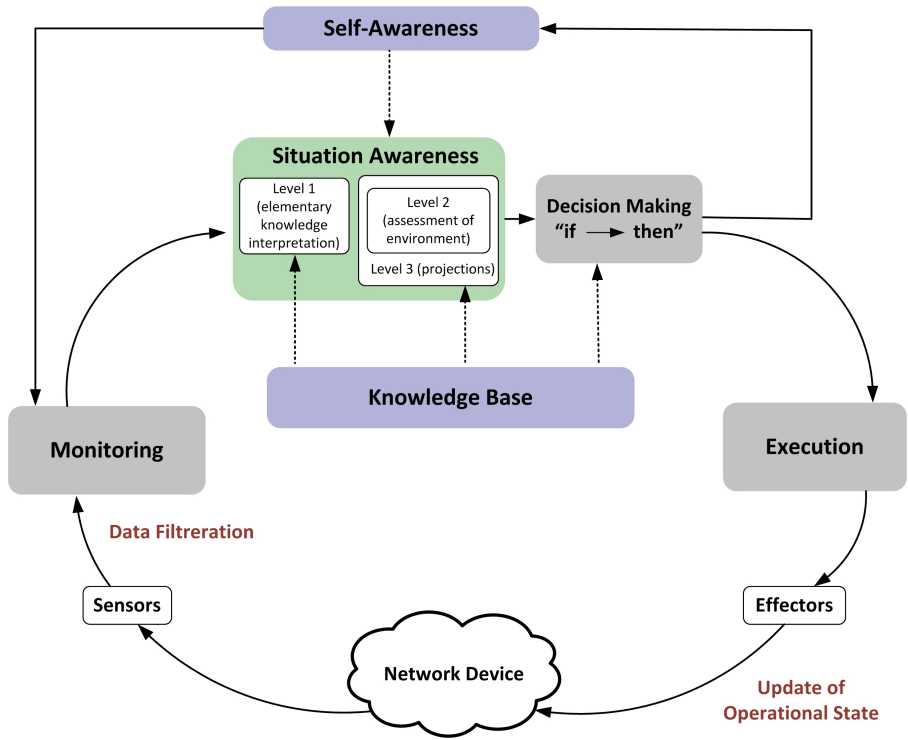}
\caption{Cognitive cycle }
\label{cycle}
\end{figure*}

\section{Cognitive Network Management}

Cognition is central to future Internet self-managed systems. The concept of self-management is derived from IBM's vision of autonomic computing, which is based on four key aspects of self-configuration, self-optimization, self-healing, and self-protection \cite{vision_AC}. This concept of autonomic computing has been further extended to autonomic networking. In order to create self-managed systems, the aspects of self-organization and self-awareness are also necessary. Autonomic operation of a system or
a network is achieved using feedback/control loops or cognitive cycles.

The cognitive cycle (shown in Fig. \ref{cycle}) can be decomposed into two parts, each operating at a different scale \cite{cognition}:
\begin{itemize}
\item
A reactive part, operating at a shorter time scale, that senses and acts directly on a network device based on pre-defined rules.

\item
A learning part, operating at a longer time scale, that exploits feedback from previous events or historical data to extract knowledge and adapt decision rules accordingly.
\end{itemize}

The cognitive cycle is also known as the $\mathbb{MDE}$ (Monitoring, Decision Making, and Execution) cycle. In a generalized form, the $\mathbb{MDE}$ cycle involves interactive feedback steps for collecting inputs from the environment and the involved elements (Monitoring), reasoning and learning based on certain algorithms and available knowledge (Decision Making), and invoking actions for achieving desired goals in the system (Execution).

A fully fledged cognitive cycle also includes more advanced aspects such as self-awareness and situation awareness. Self-awareness can be seen as the network device's view on internal and external processes, statuses, and states needed for conducting the deduction processes in the cognitive cycle. On the other hand situation awareness is the step that precedes and constitutes the foundations for decision making. To perform situation awareness, various knowledge types and sources would be consulted in different stages of interpretation. Generally, three different levels of situation awareness exist. The level 1 refers to the characterization of operational state (of a segment of a system, physical and/or logical). In level 2, an assessment of environment is carried out, based on which projections or predictions are made in level 3 for proceeding to the decision making step.

The functional entity that encompasses the cognitive cycle functionalities is referred to as the Cognitive Network Manager (CNM). The CNM, which is an autonomic entity (also referred to as the agent) and has been introduced in \cite{cognition,agent}, is capable of detecting device anomalies and network service disruptions, diagnosing root causes, reporting load information, and computing possible corrective actions whenever necessary. For more details on the CNM, the interested reader is referred to \cite{cognition}.

\section{Proposed Mechanisms}
The proposed mechanisms are presented in the form of two independent capabilities. Before discussing the details of these capabilities, it is important to describe the baseline architecture.

\subsection{Functional Architecture}
We consider a tightly-coupled cellular/Wi-Fi system wherein both networks share the same core network. We assume a two-tier architecture and use two different types of CNMs. The simple CNMs are referred to as the Network Element Cognitive Managers (NECMs) whereas, the domain CNMs are referred to as the Network Domain Cognitive Manager (NDCMs). The NECM implements the cognitive cycle at the network element level and has the local visibility of different network entities such as Wi-Fi access points, cellular base stations, etc. On the other hand, the NDCM, which is located in the core network, implements the CNM at the network level and has an end-to-end visibility of the entire network. The NDCM utilizes the cognitive capabilities to identify optimization opportunities considering the network status and the cooperation from NECMs. Apart from this, we assume another CNM in the cellular network which has the visibility of a group/cluster of base stations. This CNM is termed as the Network Configuration Cognitive Manager (NCCM).

\subsection{Capability of Dynamically Powering Down the Radio Network Equipment}
\begin{figure*}
\centering
\includegraphics [scale=0.70] {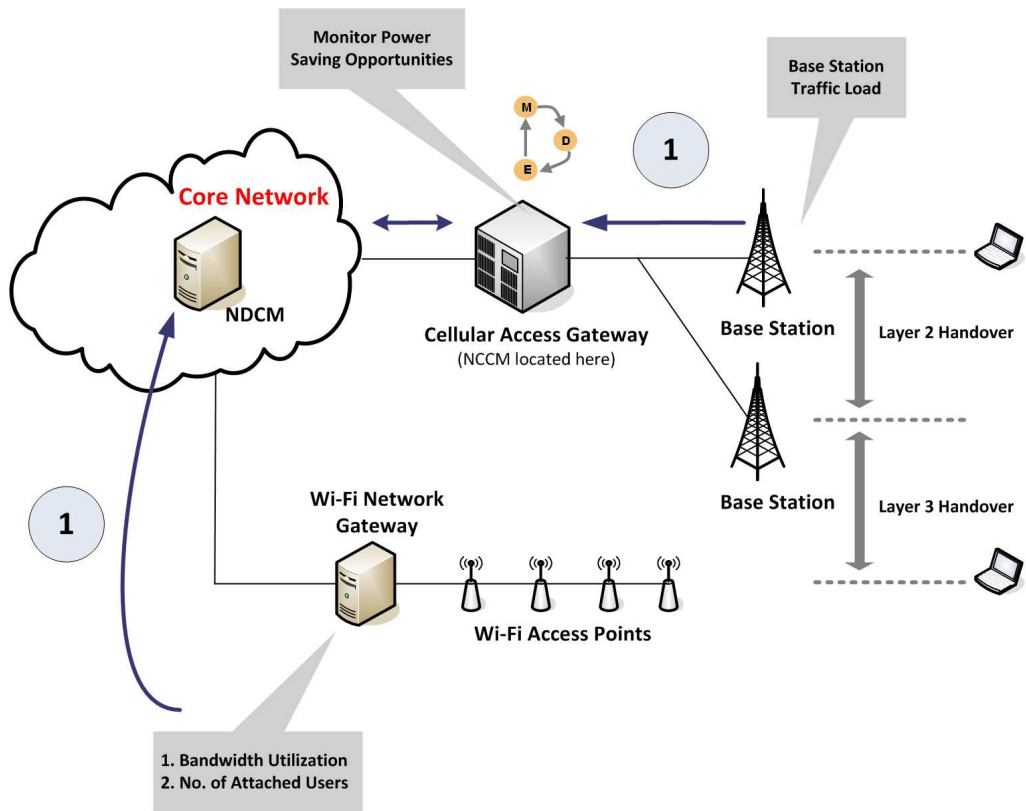}
\caption {$\mathbb{MDE}-1$ (for dynamically powering down the radio network equipment)}
\label{MDE1}
\end{figure*}

The first capability provides a network-centric and short-term solution for reducing energy
consumption, and supports legacy networks. The traffic load on cellular and Wi-Fi networks
is monitored and the two power saving solutions; sectorization switching and powering down
are opportunistically applied. For this capability, the different steps of the $\mathbb{MDE}$ cycle (termed as $\mathbb{MDE}-1$) are described as follows (also illustrated in Fig. \ref{MDE1}).

\begin{itemize}

\item \emph{Monitoring and Situation Awarenenss} -- During this phase, traffic load on cellular and Wi-Fi networks is monitored. The NECMs present in the cellular base stations monitor the load of each base station and periodically report to the NCCM. Similarly, the NECMs present in the Wi-Fi access points monitor the traffic load (in terms of bandwidth utilization and number of attached users) along with some other Quality of Service (QoS) indicators and periodically report to the NDCM via the Wi-Fi network gateway. 

The level 1 of situation awareness here refers to characterizing the operational state of base stations depending on the load i.e., if the base station is operating in omni-directional mode or sectorized mode. The level 2 of situation awareness refers to identifying potential Wi-Fi access points for opportunistically offloading users. In the final step of situation awareness, projections are made for selecting a suitable power saving solution. 

A trigger is initialized to proceed to the decision making phase once the traffic load on cellular network reaches $T_{switch}$ (threshold at which omni-directional mode is switched to tri-sectorized mode).

\item \emph{Decision Making} -- During this phase a decision is made about either removing the sectorization or powering down
a base station depending upon the capacity available on the Wi-Fi network. In the simple case, this decision would be made by the NCCM which has cluster level visibility of cellular network (in terms of base stations in the cluster and their traffic loads). To avoid the so called ping-pong effect, no decision will be made as long as the traffic load stays at $T_{switch}$.  The actual trigger points can be defined in terms of percentage, e.g. sectorization switching solution can be triggered when the traffic load falls to 90\% (or below) of $T_{switch}$. Similarly powering down solution can be triggered when the traffic load falls to 10\% (or below) of the busy hour load. In a more complicated and realistic case, where the Wi-Fi access points are overlapping between the coverage areas of multiple base stations, the algorithm will decide whether to remove sectorization from all base stations or power down one base station considering maximizing the energy savings on the cellular network.

\item \emph{Execution} -- During the execution phase, the selected base station is switched to omni-directional mode or powered down completely. In the former case, i.e., sectorization switching, no further actions are required. On the other hand, powering down a base station results in a coverage hole. In order to overcome this issue, we propose base station cooperation wherein neighboring base stations cooperate to provide coverage in the service area of powered-off base station. In our previous work \cite{pimrc_13}, we investigated different base station cooperation patterns and evaluated the outage probability for each scenario. The users attached to the powered-off base station are offloaded to the neighboring base stations or to the Wi-Fi network. In the next section, we propose a  network-centric Wi-Fi offloading solution, which is specifically designed for this capability. 

\end{itemize}

\begin{figure*}
\centering
\includegraphics [scale=0.63] {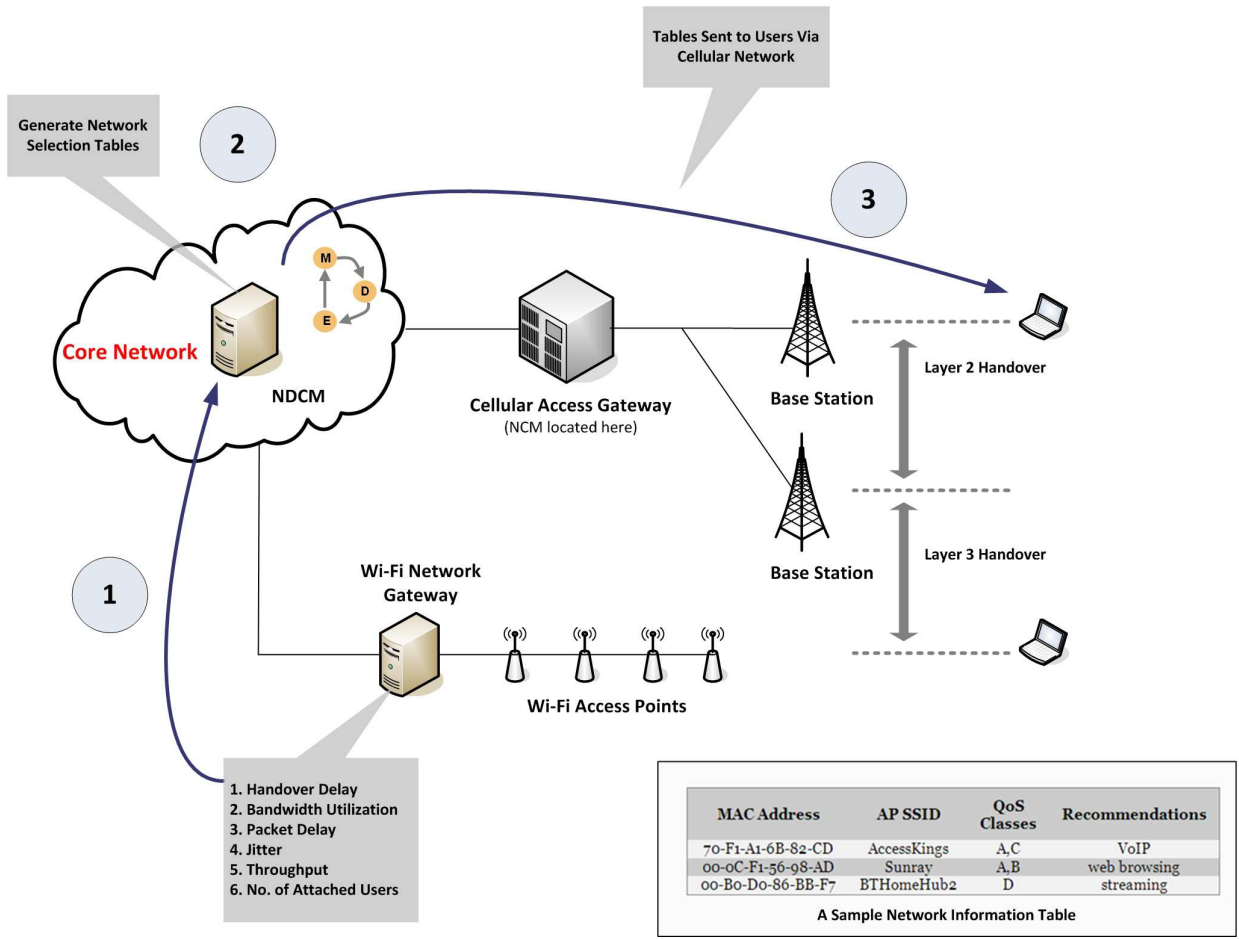}
\caption {$\mathbb{MDE}-2$ (for energy aware network selection). A sample NIT is also shown.}
\label{MDE2}
\end{figure*}

\subsection{Capability of Energy Aware Network Selection}
The second capability presents a long term and user-centric solution where users are
given more freedom and provides a future business model as well. It facilitates reduction in traffic load on the cellular network by proactively encouraging users to shift to the Wi-Fi network. The key concept is to periodically send Network Information Tables (NIT) to users, with information containing the availability of Wi-Fi access points, supported QoS classes, and recommendations for suitable services. The different steps of the  $\mathbb{MDE}$ cycle (termed as  $\mathbb{MDE}-2$) for this capability are discussed as follows (also illustrated in Fig. \ref{MDE2}).

\begin{itemize}

\item \emph{Monitoring and Situation Awarenenss} -- During this phase, traffic load on the Wi-Fi network is monitored. The NECMs monitor the bandwidth utilization, number of attached users, and handover delay associated with each Wi-Fi access point. Apart from monitoring nodal information, end-to-end throughput, packet delay and jitter are also monitored in coordination with the NDCM. The NECMs periodically send this information to the NDCM via the Wi-Fi network gateway.

The level 1 of situation awareness refers to the characterization of operational state of the Wi- Fi access points based on the information sent by the NECMs, e.g. if an access points is highly utilized or underutilized in terms of the bandwidth and the number of attached users. The level 2 refers to checking any backhaul issues and identifying the congested parts of the network. In level 3, projections are made regarding QoS classes supported by different Wi-Fi access points.

\item \emph{Decision Making} -- During the decision making phase, NITs are generated by the NDCM based on the information sent by the NECMs. As shown in Fig. \ref{MDE2}, these tables include information about the MAC address of the Wi-Fi access point, the Service Set Identifier (SSID), the supported QoS classes\footnote{e.g., 3GPP QoS classes: conversational, streaming, interactive, and background} and recommendations for different services. The classification into different QoS classes is done with the help of databases (knowledge base in the cognitive cycle) containing a range of values that map a service to different parameters such as packet delay, throughput, jitter, packet loss etc. Based on this classification, the recommendations for different services (maximum supported service per access point) are included.

\item \emph{Execution} -- During this phase, the NITs would be sent to the users. These tables would be sent periodically via the cellular network. It is important to decide the scale and size of sending these tables. Ideally the table size should be kept small in order to reduce the signalling load on the cellular network. One way is to send the tables on per cluster area basis. This is recommended as the energy saving solutions have cluster level impact. The NDCM can generate tables for the entire network. This data would be sent to the NCCM which will then filter down and broadcast the shorter versions of these tables for that specific location. One way of performing the filtering is with the help of databases containing the location of Wi-Fi access points. Alternatively, the tables can be sent on per location area basis just like the cellular paging messages. The NCCM will compare the tables received from the NDCM with the previous versions and will only send the new information when there is a change/update. Besides, the network unicasts complete tables to all terminals attaching on the network for the first time. On receiving these tables, terminals have the ability to make their own decisions as to which network will be used for a particular service. Alternatively, the operator can take a more proactive role in enforcing these tables and requiring the terminal to use alternative networks with the aim of maximizing energy savings on the cellular network by facilitating powering down of selected base stations. In the next section, we propose a user-centric Wi-Fi offloading solution, which is specifically designed for this capability.  $\blacksquare$



\end{itemize}

The two $\mathbb{MDE}$ cycles described above may co-exist and run in parallel. This would result in increased energy savings as $\mathbb{MDE}-2$ will create more opportunities for removing sectorization and powering down by facilitating reduction in traffic load on the cellular network.  Besides achieving energy savings in the cellular access network, the proposed mechanism in the form of the two $\mathbb{MDE}$ cycles has other benefits. Firstly, the proposed mechanism is generic and not tied to specific network topologies. Secondly, users are offloaded to Wi-Fi network while considering their QoS requirements. Last, but not the least, the $\mathbb{MDE}-2$ cycle provides a promising solution for  encouraging users to switch to Wi-Fi. This is particularly important for those mobile operators who want to dynamically offload traffic to Wi-Fi network but cannot force the terminals to keep both cellular and Wi-Fi interfaces simultaneously on\footnote{It is not suggested to keep both cellular and Wi-Fi interfaces simultaneously on due to significant battery usage, especially when the Wi-Fi interface is in Idle mode}. By using the NITs, received from the cellular network, users obtain information about suitable Wi-Fi access points and switch to Wi-Fi whenever possible.

\section{IFOM-based Wi-Fi Offloading Solutions}
It is important to discuss the Wi-Fi offloading technique as it is an integral part of the proposed mechanisms. Currently, three different techniques are under consideration by 3GPP that will shape offloading from cellular networks. These include Local IP Access (LIPA), Selected IP Traffic Offload (SIPTO), and IP Flow Mobility (IFOM). In this paper the main focus is on IFOM since both LIPA and SIPTO are used for offloading from macrocells to small cells (femto, pico, etc.). In \cite{IFOM}, two different techniques have been discussed for implementing flow mobility solutions. We adopt these techniques and discuss how they can be employed for offloading users to Wi-Fi networks in the proposed mechanisms to achieve energy savings. 

\subsection{IP Flow Mobility}

IP flow mobility is a recent technology that is currently being standardized in the Internet Engineering Task Force (IETF). This technology allows an operator to shift a single IP flow to a different radio access without disrupting any
ongoing communication. Consider a user connected to a cellular base station having multiple simultaneous flows (e.g., a voice call and a file download) moving into the coverage of a Wi-Fi access point (hotspot). The terminal or network, upon detection of the Wi-Fi access, decides to shift the file download on the Wi-Fi network. Once the user leaves the Wi-Fi coverage, the file
download is seamlessly shifted back to the cellular network.

%

\begin{figure*}
\centering
\subfloat[]{\label{uc_IFOM}\includegraphics [scale=0.65] {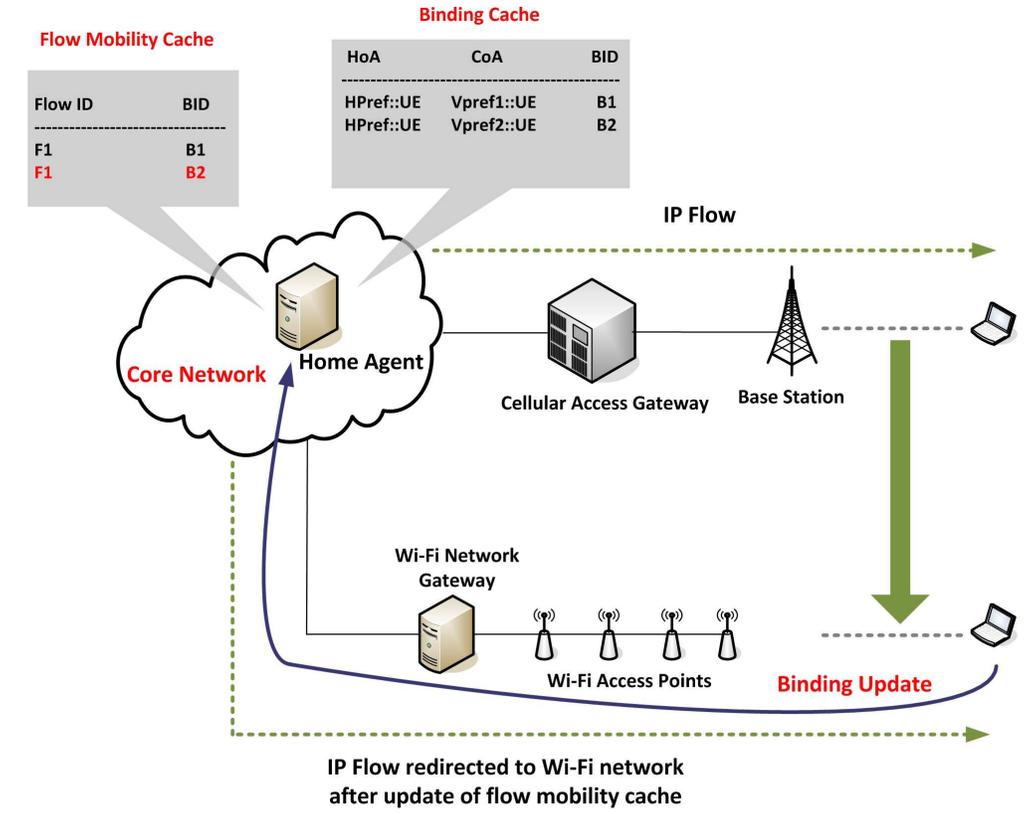}} \qquad
\subfloat[]{\label{nwc_IFOM}\includegraphics [scale=0.65]{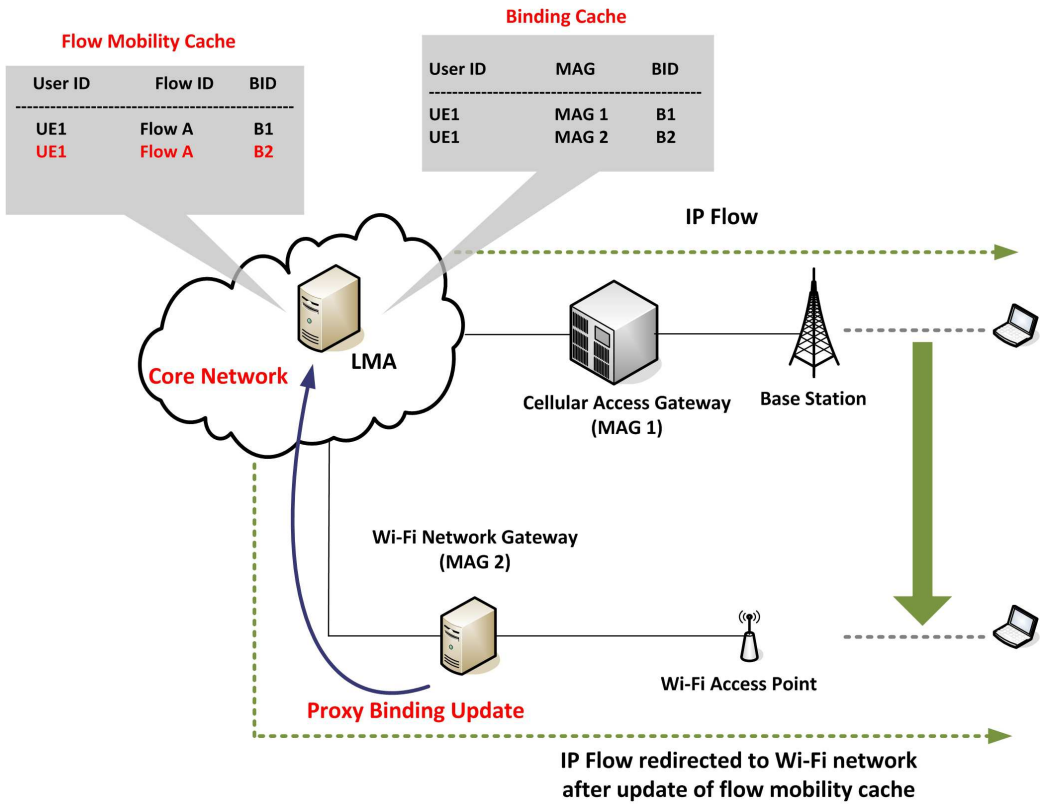}}
\caption{IFOM-based Wi-Fi offloading solutions, (a) user-centric approach, (b) network-centric approach.}
\label{IFOM_soln}
\end{figure*}

\subsection{User-centric IFOM-based Wi-Fi Offloading}
In the user-centric offloading solution, illustrated in Fig. \ref{uc_IFOM}, the user is involved in the mobility management by detecting and signalling changes to points of attachment. This solution is based on Mobile IPv6 (MIPv6) \cite{rfc3775} with some flow mobility extensions. In MIPv6, global reachability is achieved through an entity known as the home agent (HA), which anchors the permanent IP address of a user in the home network called the home address (HoA). When away from the home network, the user (mobile node) obtains a temporary IP address from the visited network called the care-of address (CoA), and informs the HA about its current location through a binding update message. The HA establishes a bi-directional tunnel to re-direct traffic to and from the user. IFOM functionality using MIPv6 requires two key extensions: (a) multiple CoA registration support, and (b) flow binding support at the HA. Multiple CoA registration allows a user to register multiple CoAs with the same HoA to achieve flow mobility. The flow binding allows a user to bind one or more IP flows to a specific CoA. 

The user-centric offloading solution provides a natural offloading mechanism for the capability of energy aware network selection ($\mathbb{MDE}-2$). As shown in Fig. \ref{uc_IFOM}, the user on receiving the NITs can switch to the Wi-Fi network, obtain a new CoA, and send a binding update message to the HA. The HA updates the binding cache and adds a new binding entry for the user with a new binding ID. Due to flow binding support, the user associates  a particular IP flow with a particular CoA. The HA maintains a flow mobility cache associating specific IP flows with  binding IDs and therefore re-directs the traffic to the user when it is on the Wi-Fi network. 


\subsection{Network-centric IFOM-based Wi-Fi Offloading}
In the network-centric offloading solution, illustrated in Fig. \ref{nwc_IFOM}, users are not involved in the mobility management and IP signalling. This solution is based on extended Proxy Mobile IPv6 (PMIPv6) \cite{rfc5213} protocol, wherein flow mobility is achieved by making the physical interface (cellular or Wi-Fi) transparent to IP and higher layers. In PMIPv6, mobility management is handled by two key entities: a Mobile Access Gateway (MAG) and a Local Mobility Anchor (LMA). The MAG performs mobility related signalling on behalf of the users attached to its access links. Typically, the MAG is the access router for the user. The LMA resides in the core network and acts as a local HA for the user. In order to achieve IFOM functionality using PMIPv6, the IETF has proposed to implement a Logical Interface (LIF) that combines several physical interfaces into a unique interface. The LIF is a software entity that hides the real physical interface implementation to IP and higher layers. It is typically implemented as part of the connection manager software of the user equipment.

The network-centric offloading solution provides a natural offloading mechanism for the capability of dynamically powering down the radio network equipment ($\mathbb{MDE}-1$). In the \emph{execution} phase, users are offloaded to Wi-Fi network if they are within the Wi-Fi coverage. Consider the scenario shown in Fig. \ref{nwc_IFOM}, wherein a user is attached to the cellular network and has an active flow through the MAG-1. When the user is offloaded to the Wi-Fi network, MAG-2 detects its attachment on the access link and carries out the necessary authentication procedure. After this MAG-2 sends a proxy binding update message to the LMA, after which the LMA updates its binding cache and creates a bi-directional tunnel with the MAG-2. Note that due to the LIF, the actual physical interface is transparent to the LMA. Next, the LMA updates its flow mobility cache and re-directs the flow to the MAG-2. Thus, the user achieves service continuity. $\blacksquare$


Note that the IFOM functionality in both user and network centric solutions also allows selected traffic flows to be offloaded to the Wi-Fi network and therefore, facilitates the application of sectorization switching solution when the available capacity on the Wi-Fi network is not sufficient for the powering down solution.

\begin{figure}
\centering
\includegraphics [scale=0.19] {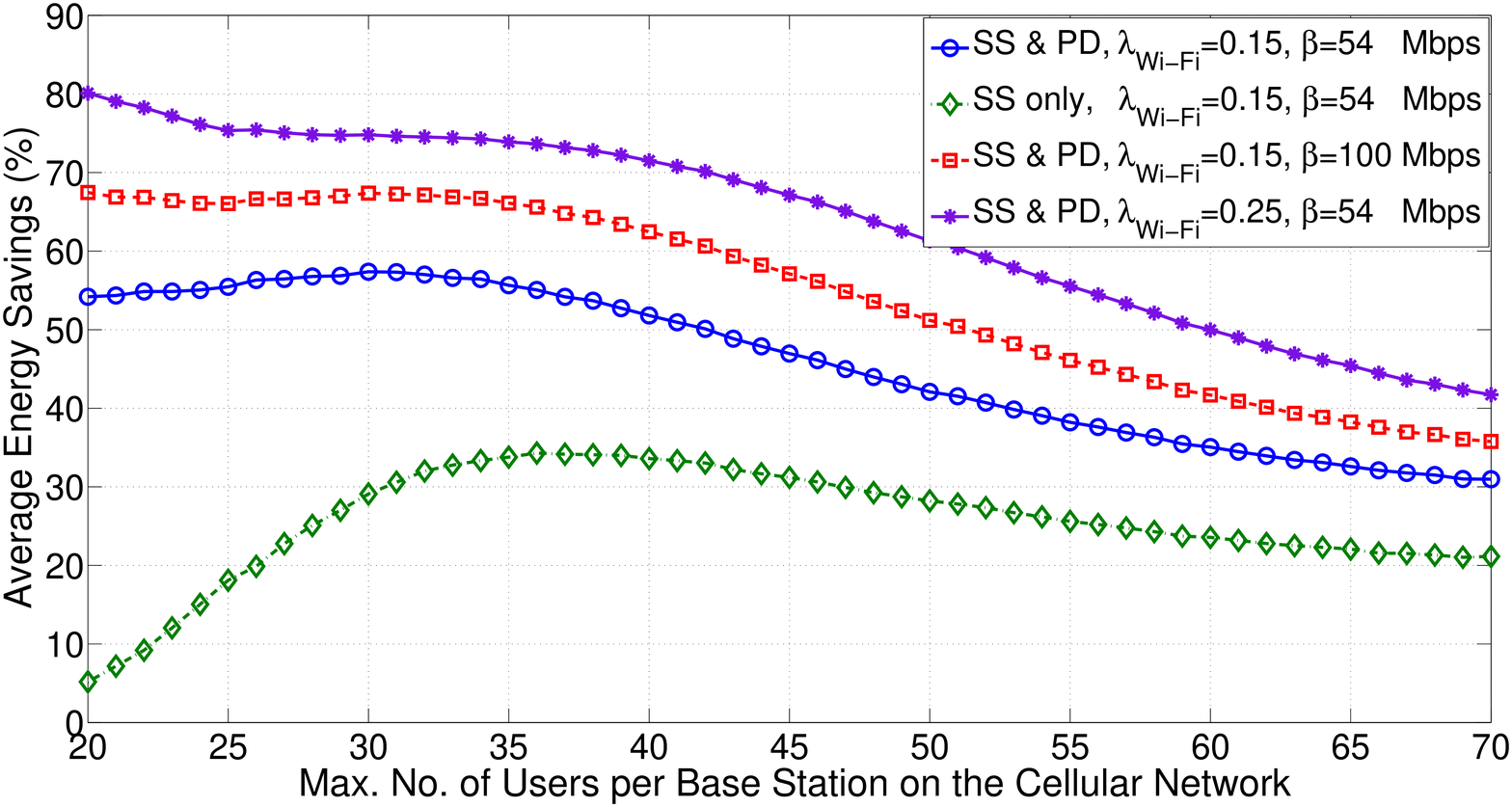}
\caption {Average energy savings for the capability of dynamically powering down the radio network equipment.Abscissa represents the mean of Poisson distributed active users. SS and PD refer to sectorization switching and powering down solutions respectively.}
\label{cap1}
\end{figure}

\section{Performance Evaluation}

Unlike \cite{aijaz_icc_12} where performance evaluation has been conducted in a single-cell scenario, we quantify the energy saving potential of the proposed energy saving concepts in realistic multi-cellular scenarios and aided by proposed mechanisms.

For the first capability of dynamically powering down the radio network equipment, we consider a $7$-cell cluster of base stations. Each base station has a coverage radius of $300$ meters wherein users are uniformly distributed with a minimum distance of $20$ meters from the base station.  The Wi-Fi access points are Poisson distributed in the overall coverage area of the cellular cluster with a density $\lambda_{Wi-Fi}$. Each Wi-Fi access point has a coverage radius of $50$ meters and a capacity of $\beta$ (in Mbps). Further, we assume that a Wi-Fi access point consumes $1\%$ of the energy consumed by a conventional cellular base station.  We assume that the number of active users on the cellular as well as the Wi-Fi networks is Poisson distributed with the mean given by the time-varying traffic distributions \cite{aijaz_icc_12} of both networks. As mentioned earlier, for the case of powering down solution, we use base station cooperation strategies \cite{pimrc_13}. As shown by the cooperation patterns therein, up to $4$ base stations can be powered off in a $7$-cell cluster to avoid outage in the total coverage area. 


The results in Fig. \ref{cap1} show the average energy savings (from-the-socket savings) over a $24$-hour period for the cellular network using the first capability. As shown by the results, combined application of both energy saving techniques yields higher energy savings than the application of a single technique. The energy savings by turning the base station off at low loads are most notable. At high loads, the primary contribution is from the sectorization switching solution which provides maximum savings of up to $35\%$. The energy savings are dependent on the capacity of the Wi-Fi access points. With more capacity per Wi-Fi access point, more users can be offloaded. Similarly, denser Wi-Fi deployments create more offloading opportunities which result in higher energy savings.


For the capability of energy aware network selection, we consider  a similar scenario as before. Besides, we consider three type of services; voice over IP (VoIP),  web-browsing, and video streaming, with delay requirements of up to $100$ms, $500$ms, and $250$ms, respectively. We adopt ON/OFF traffic models for these services with parameters given in \cite{aijaz_icc_12}.  Further, we assume that the delay on the Wi-Fi network increases exponentially between $50$ms and $600$ms when Wi-Fi utilization increases between $50\%$ and $100\%$. Each active user is assumed to have a VoIP, web-browsing, or video streaming  session with equal probability. Besides, the active user will make an offloading decision with a probability of temporal Wi-Fi coverage ($\mathcal{P}_t$), which is selected as $0.65$ for the day time ($09:00 - 18:00$) and $0.85$ for the night time.

\begin{figure}
\centering
\includegraphics [scale=0.19] {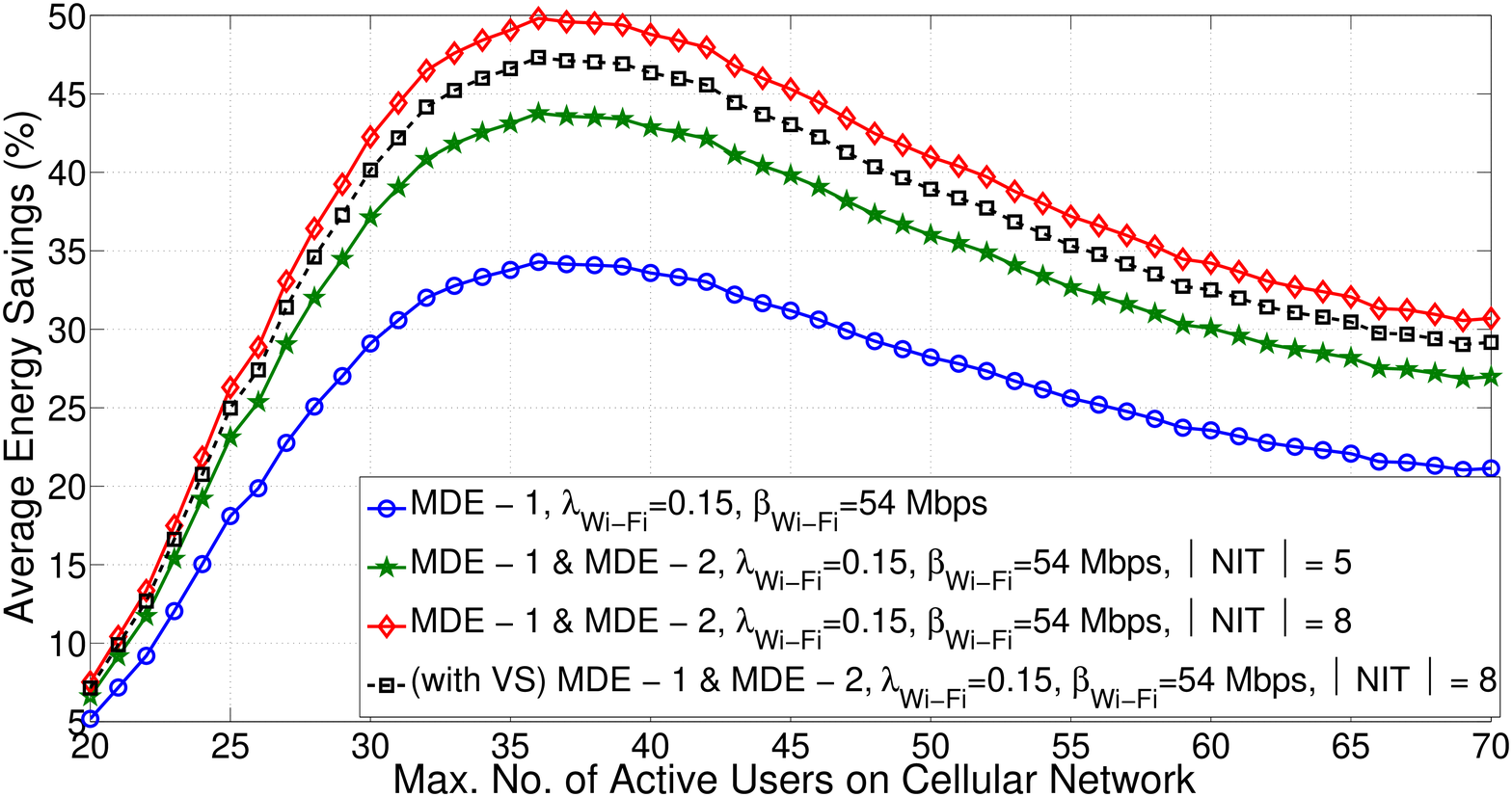}
\caption {Average energy savings for the capability of dynamically powering down the radio network equipment and the capability of energy aware network selection. Only sectorization switching solution is applied. Abscissa represents the mean of Poisson distributed active users. VS refers to video streaming.}
\label{cap2}
\end{figure}

Fig. \ref{cap2} shows the average energy savings by applying the sectorization switching solution for both capabilities. Note that running the two $\mathbb{MDE}$ cycles in parallel results in significantly higher energy savings. The energy savings are also dependent on the size of NITs. Increasing the size of NITs creates more offloading opportunities by providing more potential Wi-Fi access points for offloading. We evaluate the energy savings with and without video streaming services. Note that the energy savings reduce with video streaming flows (along with VoIP and web-browsing). This is because video streaming generates a higher traffic load per user compared to VoIP and web-browsing. 

Lastly, it is important to mention that the computational complexity of the $\mathbb{MDE}$ cycle has been evaluated experimentally in \cite{cognition}.

\section{Concluding Remarks}
Energy consumption in mobile communications is an important issue which, like in all areas of technology, must be reduced to aid the environment. Mobile network operators are increasingly offering connectivity over heterogeneous networks, blending licensed and unlicensed spectrum. Against these observations, we have proposed mechanisms for energy savings in cellular access network through opportunistic reallocation of traffic load to Wi-Fi networks. The proposed mechanisms are based on the principles of cognitive network management. Further, two different Wi-Fi offloading solutions have been proposed based on IFOM. The proposed mechanisms not only provide a promising solution for achieving energy efficiency in cellular networks, but also address some of the key challenges faced by operators in offloading traffic to Wi-Fi networks.



%



%
%
%
\section*{Acknowledgment}
The authors would like to thank Dr. Paul Pangalos, Dr. Oliver Holland, and Dr. Andrej Mihailovic of King's College London for partaking in fruitful discussions.
%
%



\bibliographystyle{IEEEtran}
\bibliography{gw}
%

%

\begin{IEEEbiography}{Adnan Aijaz}
(M'14) received the M.Sc degree in Mobile and Personal Communications and the Ph.D degree in Telecommunications Engineering from King's College London, in Oct. 2011 and Oct. 2014 respectively. Currently, he is working as a Research Associate with the Centre for Telecommunications Research, King's College London. His research interests include 5G cellular networks, machine-to-machine communications, full-duplex communications, cognitive radio networks, smart grids, and molecular communications. He has served as  symposium co-chair and TPC member for IEEE SmartGridComm'15 and IEEE VTC Spring'15, respectively. 

Prior to joining King's, he worked in cellular industry for nearly 2.5 years in the areas of network performance management, optimization, and quality assurance. He obtained the B.E. degree in Electrical (telecom) Engineering from National University of Sciences and Technology (NUST), Pakistan.
\end{IEEEbiography}

\begin{IEEEbiography}{Abdol-Hamid Aghvami}
(M'89--SM'91--F'05) is a Professor of Telecommunications Engineering at King's College London. He has published over 600 technical papers and given invited talks and courses world wide on various aspects of Personal and Mobile Radio Communications. He was Visiting Professor at NTT Radio Communication Systems Laboratories in 1990, Senior Research Fellow at BT Laboratories in 1998-1999, and was an Executive Advisor to Wireless Facilities Inc., USA, in 1996-2002. He was a member of the Board of Governors of the IEEE Communications Society in 2001-2003, was a Distinguished Lecturer of the IEEE Communications Society in 2004-2007, and has been member, Chairman, and Vice-Chairman of the technical programme and organising committees of a large number of international conferences. He is also the founder of International Symposium on Personal Indoor and Mobile Radio Communications (PIMRC); a major yearly conference attracting nearly 1000 attendees.

Professor Aghvami was awarded the IEEE Technical Committee on Personal Communications
(TCPC) Recognition Award in 2005 for his outstanding technical contributions to the
communications field, and for his service to the scientific and engineering communities.
Professor Aghvami is a Fellow of the Royal Academy of Engineering, Fellow of the IET,
Fellow of the IEEE, and in 2009 was awarded a Fellowship of the Wireless World Research
Forum in recognition of his personal contributions to the wireless world. 
\end{IEEEbiography}

%
%
%





\end{document}